\documentclass[superscriptaddress, reprint,
nofootinbib,
bibnotes,
 amsmath,amssymb,
aps,
]{revtex4-1}

\usepackage{graphicx}% Include figure files
\usepackage{dcolumn}% Align table columns on decimal point
\usepackage{bm}% bold math
\usepackage{hyperref}% add hypertext capabilities
\usepackage[mathlines]{lineno}% Enable numbering of text and display math
\usepackage{xcolor}
\definecolor{verde}{rgb}{0,0.5,0}

\begin{document}

\title{Primordial Gravitational Waves from Scalar Backreaction in Axion-SU(2) Inflation
%Strong Backreaction by Scalar Tachyonic Instabilities in Axion-SU(2) Inflation
}

\author{Mattia Cielo}
\email{mattia.cielo@unina.it}
\affiliation{INFN - Sezione di Napoli, Complesso Univ. Monte S. Angelo, I-80126 Napoli, Italy}
\affiliation{Dipartimento di Fisica ``Ettore Pancini'', Università degli studi di Napoli ``Federico II'', Complesso Univ. Monte S. Angelo, I-80126 Napoli, Italy}
\affiliation{Instituto de Física Téorica UAM/CSIC, calle Nicolás Cabrera 13-15, Cantoblanco, 28049, Madrid, Spain}

\author{Matteo Fasiello}
\email{matteo.fasiello@csic.es}
\affiliation{Instituto de Física Téorica UAM/CSIC, calle Nicolás Cabrera 13-15, Cantoblanco, 28049, Madrid, Spain}

\author{Alexandros Papageorgiou}
\email{papageorgiou.hep@gmail.com}
\affiliation{Instituto de Física Téorica UAM/CSIC, calle Nicolás Cabrera 13-15, Cantoblanco, 28049, Madrid, Spain}

\author{Ema Dimastrogiovanni}
\email{e.dimastrogiovanni@rug.nl}
\affiliation{Van Swinderen Institute for Particle Physics and Gravity, University of Groningen, Nijenborgh 4, 9747 AG Groningen, The Netherlands}

\begin{abstract}
In this work, we perform the first numerical study of strong scalar backreaction in spectator chromo-natural inflation (SCNI) in the case where the spectator sector decays during inflation. The tachyonic instability in scalar fluctuations, activated as the system crosses the $m_Q = \sqrt{2}$ threshold, amplifies perturbations and may significantly alter the background dynamics. The strong scalar backreaction regime introduces an effective quartic term in the potential for the gauge field background that rapidly drives it to zero, accelerating the axion-gauge system decay. We describe the dynamics of such decay and derive the gravitational wave spectrum for a set of benchmark parameters. Interestingly, the signal may peak at interferometer scales and lie within LISA's projected sensitivity.
\end{abstract}

\maketitle

\date{\today}% It is always \today, today,

\maketitle

%\tableofcontents

%%%%%%%%%%%%%%%%%%%%%%%%%%%%%%%%%%%%%%%%%%%%%%%%%%%%%%%%
\section{Introduction}
\label{sec:Introduction}
%%%%%%%%%%%%%%%%%%%%%%%%%%%%%%%%%%%%%%%%%%%%%%%%%%%%%%%%

Cosmic inflation provides the standard framework for understanding the primordial universe, addressing the horizon, flatness, and monopole problems while generating the observed spectrum of density perturbations~\cite{Guth:1980zm,Linde:1981mu,Albrecht:1982wi,Starobinsky:1980te, Mukhanov:1985rz}. During inflation, quantum fluctuations are stretched to macroscopic scales, seeding the large-scale structure observed today~\cite{Mukhanov:1981xt,Hawking:1971ei,Starobinsky:1980te, Mukhanov:2007zz}. While this picture is in excellent agreement with current cosmic microwave background (CMB) data, forthcoming probes will considerably sharpen our view of the inflationary era, from next-generation CMB experiments and large-scale structure surveys to pulsar timing arrays and space-based gravitational wave detectors (see e.g.~\cite{SimonsObservatory:2018koc, CMB-S4:2016ple, CMB-S4:2022ght, CMB-S4:2016ple, LSSTScience:2009jmu, EPTA:2023xxk, Lentati:2015qwp, SKA:2018ckk, LIGOScientific:2016wof, TheLIGOScientific:2016wyq, LISA:2017pwj, Maggiore:2019uih, Kawamura:2020pcg}). Together, these observations have the potential to take us from consistency tests of the inflationary paradigm towards a genuine reconstruction of the particle content and interactions that shaped the primordial universe.

Axion-like particles constitute natural inhabitants of early universe scenarios. Originally introduced to address the strong CP problem~\cite{Peccei:1977hh,Weinberg:1977ma,Wilczek:1977pj}, they enjoy an approximate shift symmetry that automatically protects them from large radiative corrections, thereby resolving the $\eta$-problem that plagues generic scalar field models during inflation~\cite{DallAgata:2018ybl,Holland:2020jdh,Nomura:2017ehb,Freese:1990rb,Adams:1992bn,Maleknejad:2011jw,Maleknejad:2011sq,Maleknejad:2013npa,Adshead:2017hnc,Adshead:2016omu,Lozanov:2018kpk,Galtsov:1991un,Wolfson:2020fqz,Seto:2007tn}. String theory compactifications generically predict multiple axions with such properties, providing strong theoretical motivation for their presence in the primordial universe~\cite{Svrcek:2006yi,Arvanitaki:2009fg}. In this broader ``axiverse'' context, axions can play different roles, acting as inflatons or as spectator fields whose interactions with other sectors imprint additional, typically scale-dependent, signatures.

Axions may couple to gauge fields through e.g. a dimension-five Chern-Simons term of the form $\lambda \phi F\tilde{F}$, where $\lambda$ denotes the coupling strength, $\phi$ the axion field, and $F$ the field-strength tensor. This interaction drives gauge field production during axion evolution, leading to rich non-linear dynamics characterized by backreaction effects on both the axion background and the spacetime geometry. In the presence of non-Abelian gauge fields \cite{Adshead:2012kp,Dimastrogiovanni:2012ew,Maleknejad:2012dt,Adshead:2013nka,Maleknejad:2013npa}, the system exhibits distinctive features, including scalar instabilities for modes inside the horizon that can significantly amplify perturbations under suitable conditions ~\cite{Dimastrogiovanni:2012ew,Adshead:2013qp}. Such axion–gauge constructions have been extensively studied as sources of chiral gravitational waves \cite{Aoki:2025uwz}, non-Gaussianity \cite{Fujita:2018vmv,Agrawal:2018mrg} and, in some regimes, as mechanisms for generating enhanced scalar fluctuations and primordial black holes, with characteristic signatures across CMB, PTA and interferometer scales \cite{Dimastrogiovanni:2024xvc}.

In this work, we focus on a spectator realization of axion–$\mathrm{SU}(2)$ dynamics, often referred to as Spectator Chromo-Natural Inflation (SCNI), \cite{Fujita:2014oba, Iarygina:2021bxq, Dimastrogiovanni:2016fuu}. In this setup, the axion sector does not drive inflation but evolves as a subdominant component in a standard inflaton background. In such scenario, the most commonly assumed fate for the axion-gauge sector is a decay before the end of inflation, which turns out to be convenient to keep non-Gaussianity in the curvature perturbation under control at CMB scales~\cite{Fujita:2018vmv,Papageorgiou:2019ecb}. It is known that, as the gauge background parameter $m_Q$ decreases below a critical value, scalar fluctuations in the $\mathrm{SU}(2)$ sector become tachyonically unstable, with exponential growth in well-defined instability bands~\cite{Dimastrogiovanni:2012ew,Adshead:2013qp}. This instability has been exploited in related contexts to generate large scalar perturbations and primordial black holes~\cite{Dimastrogiovanni:2024xvc,Dimastrogiovanni:2025snj}.

Recent progress has clarified strong backreaction dynamics in closely related axion–gauge models, particularly in regimes where tensor fluctuations of the gauge sector feed back on the background and source observable gravitational waves. However, a systematic study of \emph{scalar} backreaction in spectator axion–$\mathrm{SU}(2)$ setups has so far been missing. In particular, it is important to establish whether the scalar instability can be consistently followed into a strong backreaction regime without spoiling the spectator nature of the sector, and what imprint this dynamics leaves in the tensor sector.

In this work, we provide the first numerical study of scalar perturbations backreaction in  axion – gauge field systems. We consider an $\mathrm{SU}(2)$ gauge sector coupled to an axion-like particle via a Chern--Simons interaction, while a separate scalar field drives the inflationary expansion. Our numerical simulations capture the dynamics of both the axion field and the gauge field vacuum expectation value in the presence of scalar backreaction. We identify and characterize a novel strong scalar backreaction regime in which the backreaction terms $B_\chi^\text{BR}$ and $B_Q^\text{BR}$, sourced by tachyonically enhanced scalar fluctuations, efficiently modify the effective potential for the gauge background. This, in turn, drives the gauge vev to zero and induces a sharp spike in the particle production parameter $\xi$. We also investigate the associated phenomenology focusing on the linearly sourced stochastic gravitational wave background. We find that a gravitational wave signal at interferometer scales can be within reach of future detectors such as LISA.

The paper is organized as follows. In Sec.~\ref{sec:model} we review SCNI by introducing the axion–$\mathrm{SU}(2)$ background, deriving the scalar fluctuation equations and studying their stability. In Sec.~\ref{sec:backreaction} we derive the backreation terms in the background equations of motion and present a benchmark numerical simulation that illustrates the transition from weak to strong scalar backreaction, the associated evolution of the effective potential, and the energy budgets of the various components. In Sec.~\ref{sec:pheno} we derive the resulting gravitational wave spectrum and discuss its observational prospects at interferometer scales. We conclude in Sec.~\ref{sec:conclusions} with a summary of our findings and an outlook on future directions.

%%%%%%%%%%%%%%%%%%%%%%%%%%%%%%%%%%%%%%%%%%%%%%%%%%%%%%%%
\section{The Spectator Chromo-Natural Inflation Model}
\label{sec:model}
%%%%%%%%%%%%%%%%%%%%%%%%%%%%%%%%%%%%%%%%%%%%%%%%%%%%%%%%

We consider in this work the SCNI model originally proposed in \cite{Dimastrogiovanni:2016fuu}. In such a scenario, the axion and non-Abelian gauge  sector do not appreciably affect the dynamics of inflation at the background level nor the large-scale observables such as the tensor-to-scalar ratio. It is well known that such dynamics may instead leave observable imprints at smaller frequencies, including a GW signal at interferometer scales and PBH production. 

The field content of the model comprises the inflaton field $\phi$, a pseudoscalar axion $\chi$, and a non-Abelian gauge field $A_\mu$ coupled to the axion through a Chern-Simons interaction. 
\begin{equation}
\begin{aligned}
S = \int d^4x &\sqrt{-g} \Bigg[ \,
 \frac{M_p^2}{2} R   
   - \frac{1}{2} (\partial \phi)^2 - V(\phi) 
   - \frac{1}{2} (\partial \chi)^2   
\\
& - U(\chi)  - \frac{1}{4} F^a_{\mu\nu} F^{a\mu\nu}
  + \frac{\lambda \chi}{4f} F^a_{\mu\nu} \tilde{F}^{a\mu\nu} 
\Bigg]\;,
\end{aligned}
\end{equation}
where \(F^a_{\mu\nu} \equiv \partial_\mu A^a_\nu - \partial_\nu A^a_\mu - g \epsilon^{abc} A^b_\mu A^c_\nu\) and \(\tilde{F}\) is contracted with the fully antisymmetric tensor \(\epsilon^{\mu\nu\rho\sigma}/\sqrt{-g}\), $\lambda$ is a dimensionless coupling constant and $g$ is the gauge coupling strength. Our metric follows the mostly positive convention, as defined as $ds^2=-dt^2+a(t)^2d\vec{x}^2=a(\tau)^2\left(-d\tau^2+d\vec{x}^2\right)$ in physical and conformal time, respectively. Finally, indices $a=1,2,3$ are gauge indices, while $i=1,2,3$ corresponds to spatial rotations. 

In the SCNI scenario, the inflaton potential $V(\phi)$ is sometimes left unspecified, its role being that of accommodating a viable early acceleration phase. On the other hand, the axion potential take the usual cosine \cite{Freese:1990rb} form\footnote{For an alternative class of potentials featuring a flat plateau at large field values while approximately preserving the shift symmetry see \cite{Nomura:2017ehb,Nomura:2017zqj,Dimastrogiovanni:2025snj}.} given by: 
\begin{equation}
    U(\chi) = \mu^4 \left[1  + \cos \left( \frac{\chi}{f} \right) \right]\,,
    \label{eq:potential}
\end{equation}
where the dimensionful parameters are the axion decay constant $f$ and the energy scale $\mu$. It is well known that the gauge sector admits the following background configuration compatible with isotropy and homogeneity of the universe\cite{Maleknejad:2011sq,Maleknejad:2011jw} 
\begin{equation}
A^a_0 = 0, \quad A^a_i = \delta^a_i \, a(t) \, Q(t)\,,
\end{equation}
where the locking of the gauge and spatial indices in this diagonal configuration ensures an isotropic energy-momentum tensor. The  quantities $a(t)$ and $Q(t)$ are respectively the scale factor and a function that quantifies the strength of the gauge field background.

The background equations of motion for the axion and gauge sectors take the following form (in the absence of backreaction)
\begin{align}
  &\ddot{\chi} + 3H \dot{\chi} + U_\chi 
  + \frac{3 g \lambda}{f}\, Q^2 \bigl(\dot{Q} + H Q \bigr) = 0\label{eq:chieom}\\[6pt]
  &\ddot{Q} + 3H \dot{Q} 
  + \bigl(\dot{H} + 2 H^2 \bigr) Q 
  + g Q^2 \left( 2g Q - \frac{\lambda \dot{\chi}}{f} \right) =0 \,.
  \label{eq:eoms}
\end{align}
It is convenient to introduce the following dimensionless parameters:
\begin{equation}
    m_Q \equiv \frac{gQ}{H} \qquad \xi \equiv \frac{\lambda \dot{\chi}}{2 H f} \qquad \Lambda\equiv\frac{\lambda Q}{f}\,.
\end{equation}

Enforcing the  $\Lambda\gg 2$ and $\Lambda \gg \sqrt{3}/m_Q$ hierarchies and seeking slow-roll compatible solutions for which $\ddot{\chi}$, $\ddot{Q}$ can be neglected, we find that the background Eq.~(\ref{eq:eoms}) can be reduced to the following form
\begin{equation}
    Q\simeq \left(\frac{-f \,U_\chi}{3 g \lambda H}\right)^{1/3} \qquad \xi\simeq m_Q+1/m_Q\,.
\end{equation}

This is the well known CNI attractor originally found in \cite{Adshead:2012kp}. In the present work, we will assume this regime is a good description of the axion-gauge field dynamics at the early stages of inflation. Since we are primarily interested in the process of decay of the axion-gauge system, this attractor will be violated at some point during inflation once the axion is close to the bottom of the potential and the corresponding production parameter acquires the approximate value $m_Q\sim \sqrt{2}$.

%%%%%%%%%%%%%%%%%%%%%%%%%%%%%%%%%%%%%%%%%%%%%%%%%%%%%%%%
\subsection{Decay of the axion-gauge sector during inflation}
\label{sec:decay}
%%%%%%%%%%%%%%%%%%%%%%%%%%%%%%%%%%%%%%%%%%%%%%%%%%%%%%%%

The main aim of this work is to study the instabilities in the spectrum of scalar fluctuations during the process of decay of the axion-gauge sector. This is an important aspect of the class model of models under consideration that has not been studied comprehensively before.

It has long been known that, in order to avoid large non-Gaussian contributions to $P_\zeta$ arising from axion-gauge field dynamics at CMB scales \cite{Planck:2019kim}, the safer scenarios are those where the axion-gauge sector decays before the end of inflation. This was originally pointed out in \cite{Fujita:2018vmv} (see Appendix B therein) and verified in \cite{Papageorgiou:2019ecb} (see Appendix E therein). Indeed, for $m_Q\gtrsim 2$, the non-linear sourcing of $\zeta$ via the axion fluctuation is typically stronger than the standard vacuum fluctuations. 

A simple possibility is to consider those cases where the spectator sector decays well before the end of inflation. In that case the axion field becomes massive and the corresponding contribution to $P_\zeta$ from the axion-gauge sector decays as matter during inflation. Denoting by $N_*$ the number of e-folds between the moment the axion becomes massive and the end of inflation, the ratio of contributions to $P_\zeta$ from axion dynamics $P_{\zeta_{\chi}}$ and $P_{\zeta_{\phi}}$ evaluated at the end of inflation becomes
\begin{equation}
    P_{\zeta_{\chi}}/P_{\zeta_{\phi}}\Big|_{\rm end}={\rm e}^{-3N_*}P_{\zeta_{\chi}}/P_{\zeta_{\phi}}\Big|_{\rm decay}\,.
\end{equation}
In such a scenario, even $N_*=10$ e-folds are enough to render negligible any direct contribution to $P_\zeta$ from axion-gauge field dynamics. 

\begin{figure*}[t]
 % \centering
\includegraphics[width=1.00\columnwidth]{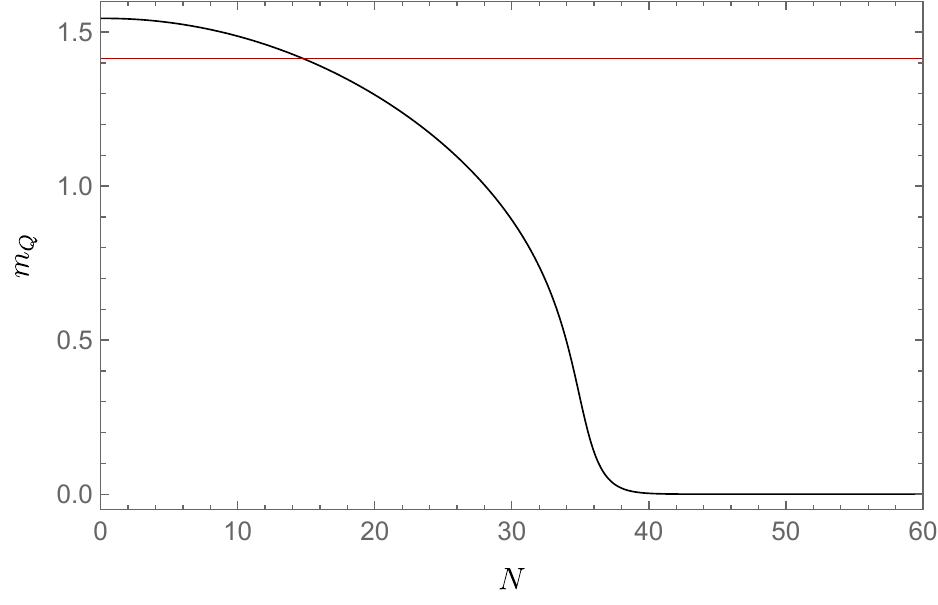}
\includegraphics[width=1.00\columnwidth]{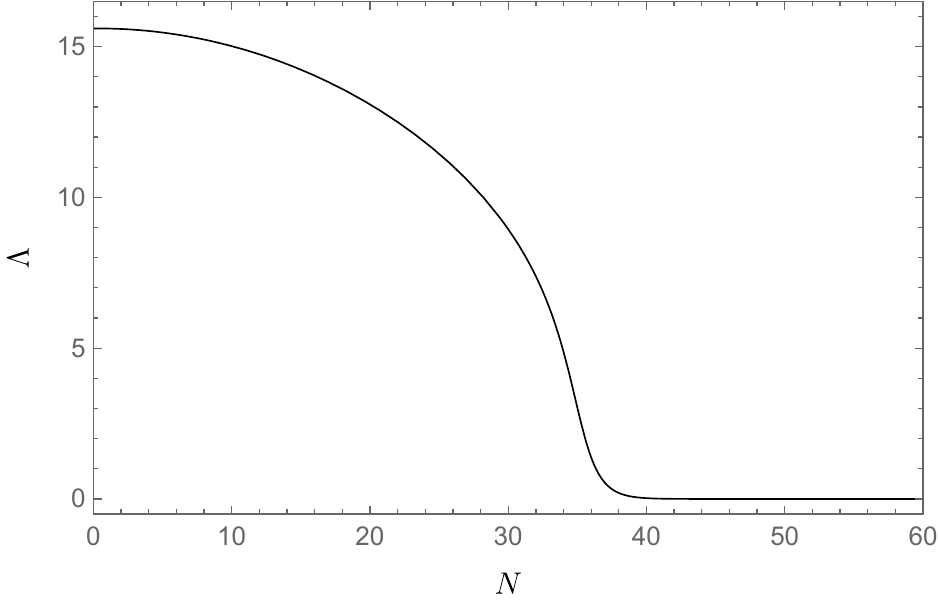}
  \caption{Typical evolution of the particle production parameters $m_Q$ and $\Lambda$ in SCNI. The maximum is obtained at the inflection point of the potential ($\chi/f=\pi/2$), and $m_Q$ reaches zero when the axion field value reaches the minimum ($\chi/f=\pi$). The red line denotes the onset of the scalar instability $m_Q=\sqrt{2}$. The parameters chosen for this example are $f=3.75\cdot 10^{-2}\;M_p$, $g=4\cdot 10^{-3}$, $\lambda=90$, $\mu=1.5\cdot 10^{-3}\;{M_p}$ and $H=1.69\cdot 10^{-5}\;M_p$. The number of e-folds is defined as $N\equiv\log(a)$ with $a=1$ at the start of the simulation.}
  \label{fig:mQ}
\end{figure*}

Studying the decay of the axion-gauge sector requires one to capture the continuous evolution of the $m_Q$ parameter from values above unity all the way to zero, see Fig.~\ref{fig:mQ}. One crucial aspect in the evolution is the presence of a well known scalar fluctuations-instability, triggered once $m_Q$ dips below $\sqrt{2}$. This threshold was originally identified in \cite{Dimastrogiovanni:2012ew} and verified in \cite{Adshead:2013nka}. Incidentally, the same instability can be used to efficiently produce primordial black holes  \cite{Dimastrogiovanni:2024xvc,Dimastrogiovanni:2025snj}.

%%%%%%%%%%%%%%%%%%%%%%%%%%%%%%%%%%%%%%%%%%%%%%%%%%%%%%%%
%%%%%%%%%%%%%%%%%%%%%%%%%%%%%%%%%%%%%%%%%%%%%%%%%%%%%%%%

We are after the effect of the scalar instability on the dynamics of the axion-gauge sector, including both background (decay) and fluctuations. In contradistinction to previous works on tensor backreaction \cite{Iarygina:2023mtj,Dimastrogiovanni:2024xvc,Brandenburg:2024awd,Dimastrogiovanni:2025snj}, our focus here is especially on the backreaction of scalar fluctuations. We therefore centre our analysis on regions of the parameter space where tensor fluctuations never become sufficiently large to significantly backreact on the axion-gauge field background. This often corresponds to choosing a sufficiently small initial value for $m_Q$. 

%%%%%%%%%%%%%%%%%%%%%%%%%%%%%%%%%%%%%%%%%%%%%%%%%%%%%%%%
\subsection{Linearized system of scalar fluctuations-stability analysis}
\label{sec:scalar}
%%%%%%%%%%%%%%%%%%%%%%%%%%%%%%%%%%%%%%%%%%%%%%%%%%%%%%%%

We focus now on the dynamics of scalar perturbations at the linearized level and study the emergence of the tachyonic instability  \cite{Dimastrogiovanni:2012ew}. This instability can be understood through the JWKB analysis of the mode functions, as detailed in \cite{Adshead:2013nka} and Appendix A of \cite{Dimastrogiovanni:2025snj}. For completeness, we re-derive here some of the main results on the tachyonic instability and elucidate its properties analytically. The next section is devoted to a numerical study.

We decompose the non-Abelian gauge field into fluctuations, focusing here on the scalar degrees of freedom, \cite{Dimastrogiovanni:2024xvc, Papageorgiou:2018rfx, Papageorgiou:2019ecb}. Our analysis follows the notation and gauge choice of \cite{Adshead:2013qp}.

\begin{align}
 \delta A^1_{\mu} &= \bigl(0,\, \Phi - Z,\, \chi_3,\, 0 \bigr)\,, \\
 \delta A^2_{\mu} &= \bigl(0,\,-\chi_3,\, \Phi - Z,\, 0 \bigr)\,, \\
 \delta A^3_{\mu} &= \bigl(\delta A^3_{0},\, 0,\,0,\, \Phi + 2Z \bigr)\,,
\end{align}
where $\delta A^3_{0}$ is a non-dynamical mode that can be integrated out of the 
equations, and
\begin{equation}
 \chi_3 = -\,\partial_z\frac{2 Z + \Phi}{2\, g\, a\, Q}\,.
\end{equation}
is the gauge-fixing condition. The constraint imposed by the non-dynamical mode $\delta A^3_{0}$ takes the form
\begin{equation}
    \delta A^3_0=-\frac{4 g a^2 \left(H Q+\dot{Q}\right)\chi_3+g a^2 Q^2 \frac{\lambda}{f}(-ik)\delta\chi}{k^2+2g^2 a^2 Q^2}\,.
\end{equation}

Upon integrating out the non-dynamical mode and gauge-fixing, we are left with three degrees of freedom. We introduce the canonically normalized field $\hat{X}$
\begin{equation}
 \hat{X} \equiv a\, \delta \chi\,,
\end{equation}
and mix the gauge field fluctuations to obtain their canonically normalized form, $\hat{Z}$ and $\hat{\phi}$,
\begin{equation}
 \hat{Z} \equiv \sqrt{2}\,\bigl(Z - \Phi \bigr)\,,
 \qquad
 \hat{\Phi} \equiv 
 \sqrt{ 2 + \frac{ k^2}{g^2 a^2 Q^2}}
 \left(
   \frac{\Phi}{\sqrt{2}} + \sqrt{2}\, Z
 \right)\,.
\end{equation}

The resulting canonically normalized quadratic Lagrangian is too long to reproduce here. Instead, we quote the final equations of motion in Appendix \ref{app:eoms}. 

If a tachyonic instability in the scalar fluctuations is present, typically it becomes significant only if it occurs deep inside the horizon. As a result, we perform an expansion of the system of Eqs.~(\ref{eq:scalarpertX}),(\ref{eq:scalarpertPsi}),(\ref{eq:scalarpertZ}) with respect to $\Lambda/x$ where $x\equiv-k\tau$. The reason we disregard terms proportional to $1/x$ is that the existence of a SCNI attractor requires $\Lambda\gg2$ and hence terms proportional to $\Lambda$ will certainly be larger than terms without it.

Deep inside the horizon the system reduces to 
\begin{align}
\hat{X}''&+\left(1+\frac{\Lambda^2 m_Q^2}{x^2}\right)\hat{X}+\frac{\sqrt{2}\Lambda}{x}\hat{\Phi}+\frac{\sqrt{2}\Lambda m_Q}{x}\hat{Z}'\simeq 0\nonumber\\
\hat{\Phi}''&+\hat{\Phi}+\frac{\sqrt{2}\Lambda}{x}\hat{X}\simeq 0\nonumber\\
\hat{Z}''&+\hat{Z}-\frac{\sqrt{2}\Lambda m_Q}{x}\hat{X}'\simeq 0\,.
\label{eq:deep}
\end{align}

To solve the system above we look for JWKB solutions of the type
\begin{equation}
    \hat{X}=A(x) {\rm e}^{i S(x)}\qquad \hat{\Phi}=B(x){\rm e}^{i S(x)}\qquad \hat{Z}=C(x){\rm e}^{i S(x)}\;.
\end{equation}
we then rewrite the resulting system (\ref{eq:deep}) as a matrix equation with respect to vector $(A,B,C)$ and impose that the determinant of the system vanishes. This results in three eigenfrequencies

\begin{align}
    &S'(x)=1\nonumber\\
    &S'(x)^2=1+\frac{3m_Q^2 \Lambda^2}{2x^2}\pm\frac{\Lambda\sqrt{8\left(m_Q^2+1\right)x^2+9 m_Q^4\Lambda^2}}{2x^2}\,.
    \label{eq:freq}
\end{align}
The first solution is physically uninteresting as it represents the exchange of energy among the gauge field fluctuations leaving the energy of the axion fluctuation unaffected. On the other hand, the other two frequencies represent a slow and fast mode, which allows for energy to be exchanged among all three scalar fluctuations. Paying particular attention to the slow mode, we see that it is possible for its frequency squared to turn negative inside the horizon. This can easily be seen by setting the lower formula of (\ref{eq:freq}) to zero and solving for $x$. We find
\begin{equation}
x_{\rm inst}=\sqrt{2-m_Q^2}\Lambda
\label{eq:tachyonic}
\end{equation}
The fact that such an $x$ exists and can be realized for reasonable values of the particle production parameter $m_Q<\sqrt{2}$ is evidence of the existence of a tachyonic instability. This tachyonic instability will also occur deep inside the horizon in a manner that is controlled by $\Lambda$ since it is proportional to $x_{\rm inst}$. Effectively, the value of $m_Q$ acts as a switch that activates the instability once it drops below $m_Q\simeq\sqrt{2}$ while $\Lambda$ modulates the strength of the instability since it controls how far inside the horizon it occurs. We can go a step further and compute the effective frequency of the instability for $1\lesssim x \lesssim x_{\rm inst}$
\begin{equation}
    S'(x)=\Omega_{\rm slow}=\pm \sqrt{\frac{m_Q^2-2}{3m_Q^2}}
\end{equation}
which is a constant (i.e. x independent). With these analytic results \cite{Adshead:2013nka} available to guide our intuition, we can proceed to numerically study the physics of this instability, we do so in the next section.

%%%%%%%%%%%%%%%%%%%%%%%%%%%%%%%%%%%%%%%%%%%%%%%%%%%%%%%%
\section{Linear strong backreaction}
\label{sec:backreaction}
%%%%%%%%%%%%%%%%%%%%%%%%%%%%%%%%%%%%%%%%%%%%%%%%%%%%%%%%

By taking the variation of the action, including the terms quadratic in fluctuations, with respect to the ALP and the gauge field backgrounds, we can obtain the following equations of motion (in cosmic time) 
\begin{align}
  &\ddot{\chi} + 3H \dot{\chi} + U_\chi 
  + \frac{3 g \lambda}{f}\, Q^2 \bigl(\dot{Q} + H Q \bigr) 
  + B_\chi^{\rm BR} = 0 \label{eq:Chieq} \\[6pt]
  &\ddot{Q} + 3H \dot{Q} 
  + \bigl(\dot{H} + 2 H^2 \bigr) Q 
  + g Q^2 \left( 2g Q - \frac{\lambda \dot{\chi}}{f} \right) 
  + B_Q^{\rm BR} =0 
  \label{eq:Qeq}
\end{align}
where the terms $B_\chi^{\rm BR}$ and $B_Q^{\rm BR}$ account for the backreaction effects due to the enhanced scalar fluctuations and take the form \cite{Dimastrogiovanni:2024xvc}
\begin{align}
 B^{\rm BR}_\chi &= 
   \frac{V^{(3)}(\chi)}{2 a^{2}} 
   \int \frac{d^{3}k}{(2\pi)^{3}} \,
   \bigl| \hat{X}(\tau, k) \bigr|^{2} \,,
   \label{eq:Bchi} \\[8pt]
 B^{\rm BR}_Q &= 
   \frac{2 g H \Lambda^{2} m_{Q}}{3 a^{2}}
   \int \frac{d^{3}k}{(2\pi)^{3}} \,
   \frac{k^{2}\left( k^{2} + a^{2} H^{2} m_{Q}^{2} \right)}{\left(k^{2} + 2a^{2} H^{2} m_{Q}^{2}\right)^{2}}
   \bigl| \hat{X}(\tau, k) \bigr|^{2} \,,
   \label{eq:BQ}
\end{align}
where we indicated with $V^{(3)}(\chi)$ the third derivative of the potential with respect to the field variable $\chi$. Note that in the above expressions, we only retain the axion fluctuations and disregard the scalar fluctuations of the gauge field. In so doing, we are implicitly assuming that the fluctuations of the axion will have a greater impact on the background dynamics in than the gauge field fluctuations. This choice is justified by the fact that the total energy of the axion is greater than that of the gauge field in all examples in this study. It is also supported by the fact that gauge field fluctuations typically acquire smaller values outside the horizon compared to the scalar fluctuation (see App.~\ref{app:initial}). Considering the full backreaction of all three scalar degrees of freedom is significantly more cumbersome, as in this case, the backreaction expression contains hundreds of terms. For this reason, we leave it to future work. 

The integrals in (\ref{eq:Bchi}) and (\ref{eq:BQ}) have a natural time-dependent ultraviolet cutoff \cite{Dimastrogiovanni:2025snj}, which is the greatest possible wave number that has become tachyonic up to a particular moment. This is given by the analytic expression (\ref{eq:tachyonic}) which, when rearranged and expressed in terms of comoving wavenumber, becomes
\begin{equation}
k_{\rm UV}(t)=\max_{t'\le t}\!\left[
 \frac{a(t')\,\lambda\, Q(t')\sqrt{2H^{2}-g^{2}Q(t')^{2}}}{f}
\right].
\end{equation}

The infrared cutoff is usually taken to be the smallest wavenumber that has ever experienced a tachyonic enhancement. Since the tachyonic instability is switched on when $m_Q<\sqrt{2}$ we can write
\begin{equation}
    k_{\rm IR}= a(t') H\Big|_{m_Q(t')=\sqrt{2}}
\end{equation}
which for every example will be a unique comoving wavenumber.

\begin{figure*}[t]
 % \centering
\includegraphics[width=1.00\columnwidth]{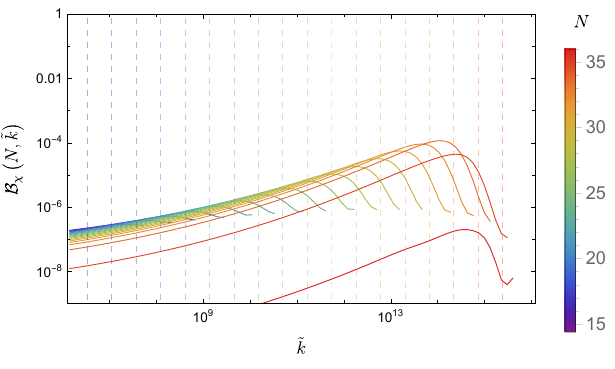}
\includegraphics[width=1.00\columnwidth]{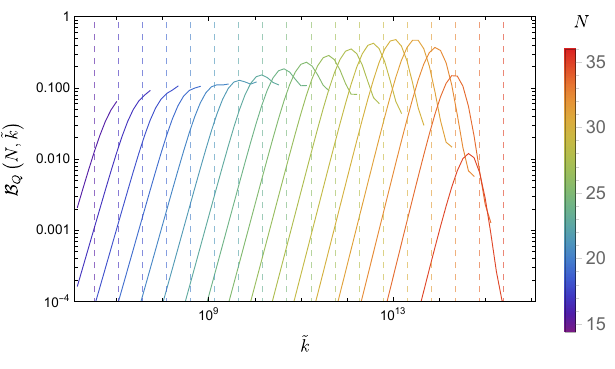}
  \caption{The fractional spectral backreaction defined in eqs.(\ref{eq:spectralbckchi}) and (\ref{eq:spectralbckQ}). The parameters chosen in this example are the ones listed in the caption of Fig.~\ref{fig:mQ}. The backreaction is well peaked both in momentum space as well as in time. The lines are only displayed between $k_{\rm IR}$ and $k_{\rm UV}$ as defined in the main text. The colored vertical dashed lines indicate the horizon crossing momentum at each snapshot in time. Note how the integral is typically dominated by scales which cross the horizon at any given time.}
  \label{fig:backreaction}
\end{figure*}

In order to get a better grasp of the properties of backreaction, it is instructive to define the \textit{fractional spectral backreaction} in the following way
\begin{align}
    &\frac{B_\chi}{U_\chi}\equiv\int d\ln\tilde{k}\; {\cal B}_\chi\left(N,\tilde{k}\right), \label{eq:spectralbckchi}\\
    &\frac{B_Q}{2 H^2 Q 
  + g Q^2 \left( 2g Q - \frac{\lambda \dot{\chi}}{f} \right)}\equiv\int d\ln\tilde{k}\; {\cal B}_Q\left(N,\tilde{k}\right)\;,
    \label{eq:spectralbckQ}
\end{align}

where $\tilde{k}\equiv k/(\sqrt{3}H)$. This recasting of the backreaction into a spectrum allows to inspect whether the backreaction is a well defined peaked integral in comoving momentum at all relevant moments in time. Additionally, the normalization factors above have been chosen strategically so that when the spectrum becomes of ${\cal O}(1)$ it is clear that backreaction turns from weak to strong. We display in Fig.~\ref{fig:backreaction} the fractional spectral backreaction for the example considered in Fig.~\ref{fig:mQ}. Note how the backreaction integrand maintains a well-defined peaked structure due to the tachyonically enhanced modes at all times. Additionally, one may notice how the particle production happens just before the axion reaches the minimum of the potential, after which the backreaction drops off. 

%%%%%%%%%%%%%%%%%%%%%%%%%%%%%%%%%%%%%%%%%%%%%%%%%%%%%%%%
%%%%%%%%%%%%%%%%%%%%%%%%%%%%%%%%%%%%%%%%%%%%%%%%%%%%%%%%

The plots in Fig.~\ref{fig:backreaction} has been obtained in the absence of the backreaction terms, an approximation not always valid, given that there are times at which the backreaction integrant is close to ${\cal O}(1)$. In the next section, we switch on strong backreaction and study how this alters the dynamics of the axion-gauge sectors.

%%%%%%%%%%%%%%%%%%%%%%%%%%%%%%%%%%%%%%%%%%%%%%%%%%%%%%%%
\subsection{Qualitative assessment of the impact of strong scalar backreaction.}
\label{sec:backreaction-potential}
%%%%%%%%%%%%%%%%%%%%%%%%%%%%%%%%%%%%%%%%%%%%%%%%%%%%%%%%

As is well known in the literature (see e.g. \cite{Papageorgiou:2018rfx}), as long as the hierarchy $\Lambda\gg\sqrt{2}$ is in place (necessary for the existence of a CNI vacuum at early times), the parameter $f$ drops out of the background equations:  the system evolves in a manner that depends only on one combination of parameters, namely $c\equiv \mu^4/(g \lambda H)$. This implies that, as long as we keep the initial condition $\chi_{\rm in}$, and parameter $c$ constant, the background evolution will remain invariant and the axion-gauge system will evolve identically at the background level regardless of the individual choices of parameter values. Note that changing $f$ always results in a change in $\Lambda$ so that the requirement of a non-zero CNI VEV identifies an upper limit on $f$, $f_{\rm max}$, for any given run. 

On the other hand, the parameters $f$ and $\Lambda$ have a dramatic effect on the scalar fluctuations that we study in this work. In fact, $f$ is the quantity that controls the size of the scalar backreaction on the background.  For any given set of initial conditions and parameter $c$ choice, values close to $f_{\rm max}$ typically lead to a suppression of the time-dependent $\Lambda$ and so a mild tachyonic instability that excites scalar fluctuations only weakly. Such large $f$ values are worth exploring in cases where the instability needs to be ``tamed" and the scalar backreaction avoided. Complementarily, when progressively decreasing $f$ we observe a greater and greater production of scalar fluctuations leading to the system entering the strong (scalar) backreaction regime.

In examples where strong scalar backreaction is triggered, we were able to identify two possible scenarios, again depending on the value of $f$. For values  $f_{\rm crit}<f< f_{\rm max}$, strong backreaction occurs in a way that only the $Q$ equation of motion is impacted by the $B_Q$ term while $\chi$ evolves non-trivially with $Q$ acting as its effective potential. In this configuration, $B_\chi$ never significantly impacts the evolution of $\chi$ directly. We will focus precisely on this case in the phenomenological section below.

When exploring values in the $f_{\rm crit}>f$ range we find explosive production of scalar fluctuations that rapidly activate both backreaction terms $B_\chi$ and $B_Q$. In this regime, we encounter numerical difficulties associated with the choice of potential. 
Specifically, if the term $B_\chi$ grows rapidly during inflation, using (\ref{eq:potential}) and (\ref{eq:Bchi}), the slope of the effective potential of the scalar field can be written as 
\begin{equation}
    dV_{\rm eff}(\chi)/d\chi=-\frac{\mu^4}{f}\sin\left(\frac{\chi}{f}\right)\left[1-\alpha(t)\right]\,,
\end{equation}
where
\begin{equation}
    \alpha(t)\equiv\frac{1}{2f^2a^2}\int \frac{d^3k}{(2\pi)^3} |\hat{X}(t,k)|^2\,.
\end{equation}
The fact that $\alpha(t)$ is positive definite and rapidly increasing implies a sudden reversal of the potential where the maxima flip to minima and vice-versa. This reversal can, of course, be traced back to the sinusoidal potential, since its first and third derivatives have the same functional form with opposite signs. This causes instabilities in our numerical code so that we do not pursue the $f_{\rm crit}>f$ regime at this stage. One ought to add that
the potential exact ``reversal'' can be prevented by including, in our expressions for scalar backreaction (\ref{eq:Bchi}), higher order terms proportional to the scalar bispectrum and trispectrum; we leave this direction to future work and focus here on the cases for which $B_\chi$ remains safely subdominant and  $B_Q$ is the main source of backreaction. 

\begin{figure}[t]
 % \centering
\includegraphics[width=1.02\columnwidth]{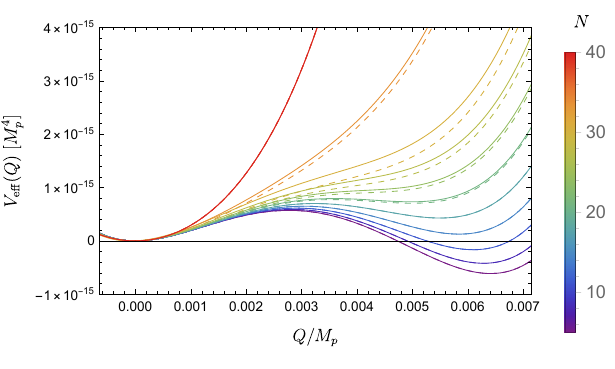}
  \caption{Time evolution of the effective potential for the gauge field background $Q$. Dashed lines correspond to the absence of scalar backreaction, whereas solid lines take into account the effective potential induced by the $B_Q$ term in the equation of motion. The shape of the potential changes over time from purple to red, and the model parameters were chosen to match the ones reported in the caption of Fig.~\ref{fig:mQ}.}
  \label{fig:effective-potential}
\end{figure}

Within the parameter range $f_{\rm max}<f<f_{\rm crit}$, it is possible to gain an intuitive understanding of the impact of backreaction on the axion-gauge background system. Upon observing that  $B_Q\propto Q^3$, when viewed as a contribution to the effective potential it is clear that the backreaction term acts as an additional quartic contribution that is centred around $Q=0$. The total effective potential can be written as
\begin{equation}
 V_{\rm eff}(Q)\equiv H^2 Q^2  - \frac{\lambda\,g \,\dot{\chi}}{3f} Q^3 +\left(\frac{1}{2}g^2+\beta(t)\right) Q^4 \;,
\end{equation}
where
\begin{equation}
    \beta(t)\equiv \frac{g^2 \lambda^{2}}{6 f^2 a^{2}}
   \int \frac{d^{3}k}{(2\pi)^{3}} \,
   \frac{k^{2}\left( k^{2} + a^{2} H^{2} m_{Q}^{2} \right)}{\left(k^{2} + 2a^{2} H^{2} m_{Q}^{2}\right)^{2}}
   \bigl| \hat{X}(\tau, k) \bigr|^{2} \,.
\end{equation}
Given that $\beta(t)$ is a positive definite, rapidly increasing function of time, the presence of backreaction acts as an external force that rapidly drives the gauge field background $Q$ to zero. This is shown in Fig.~\ref{fig:effective-potential} where the effective potential is plotted both neglecting and accounting for backreaction. The parameter values in Fig.~\ref{fig:effective-potential}  match those in Fig.~\ref{fig:mQ}, hence the overall impact is rather mild by construction. In the example we present next, the parameters are chosen so that the backreaction-induced quartic term becomes dominant, driving the gauge field background to zero rapidly. 

The rapid evolution of the gauge field background towards zero has a significant effect on the axion background since, in the CNI setup, the axion background $\chi$ is driven by the gauge field background $Q$. This effect manifests itself as a short ``spike" in the axion velocity, which copiously produces GWs just before the axion-gauge system decays. The overall size of the spike is dependent on how close the value of $f$ is to $f_{\rm crit}$, with the larger spikes occurring for values closer to $f_{\rm crit}$.

\begin{figure*}[t]
  \centering
  \includegraphics[width=\columnwidth]{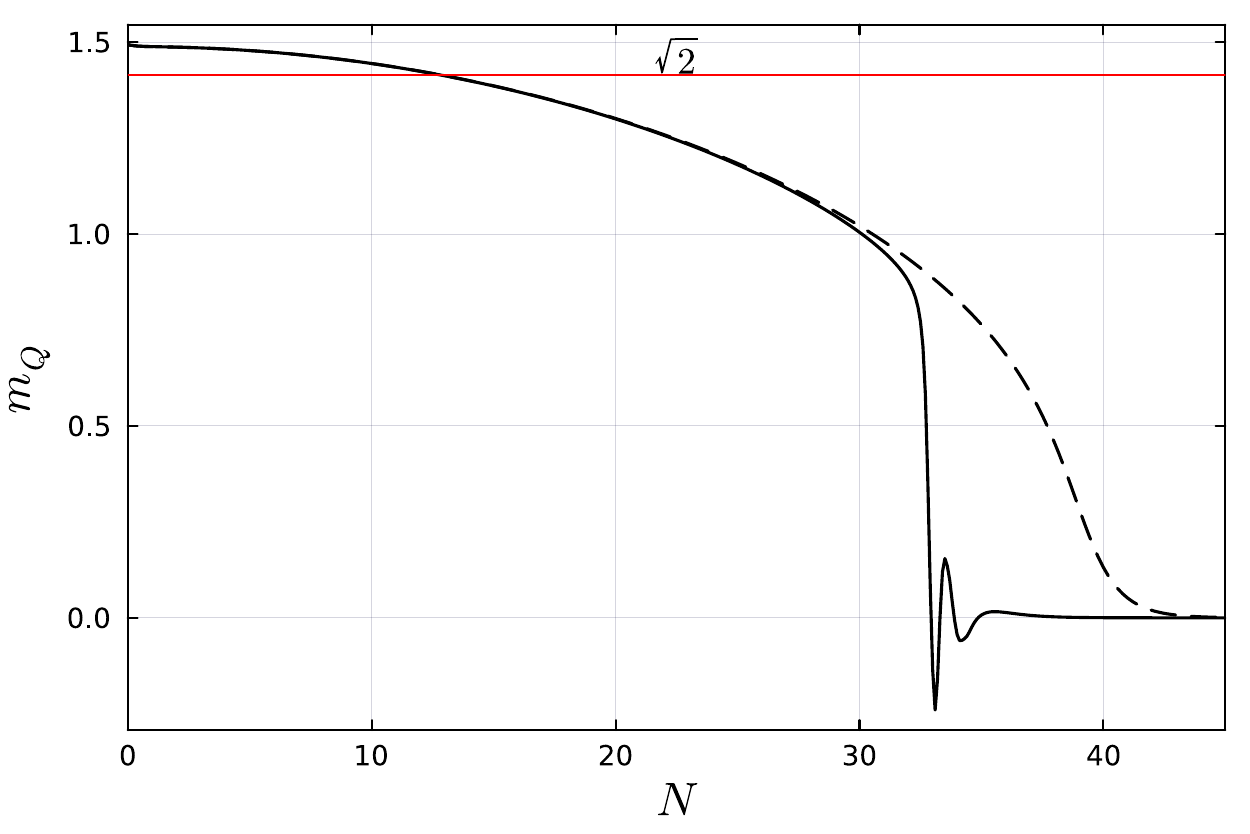}
  \includegraphics[width=\columnwidth]{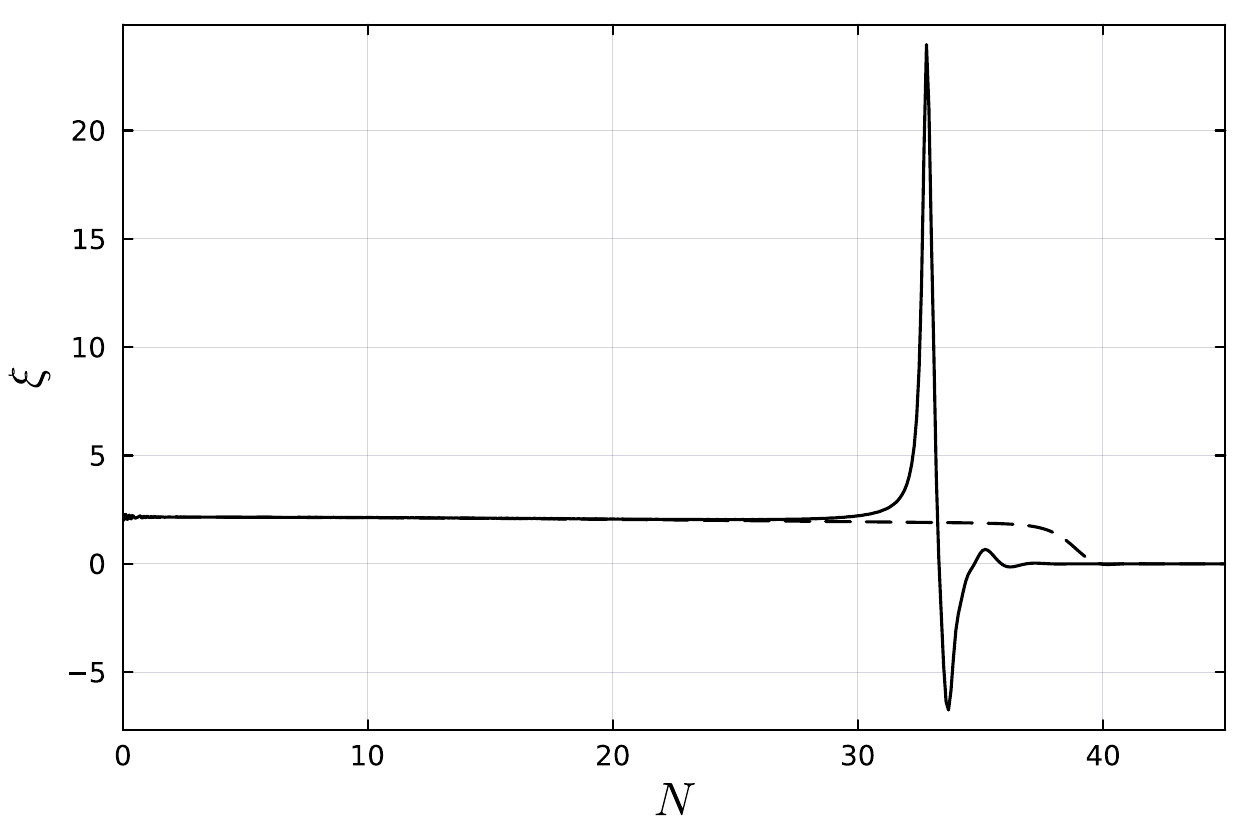}
  \caption{Particle production parameter $m_Q$ (left panel) and $\xi$ (right panel) with (solid) and without (dashed) the backreaction contribution. We use for our numerical simulation the fiducial values listed in Eq.~\ref{eq:fiducial}.}
  \label{fig:fiducial}
\end{figure*}

Let us now move to a concrete example with a parameter choice close to $f_{\rm crit}$. Our fiducial values are 
\begin{align}
    &\lambda = 100, \quad f=3.847\cdot 10^{-2}\;M_p,\quad g=4\cdot 10^{-3}\nonumber\\
    & \mu =1.5\cdot 10^{-3}\; M_p,\quad H=1.69 \cdot 10^{-5}\; M_p,\quad \chi_{\rm in}=\frac{\pi f}{2}\;.\nonumber\\
    &
    \label{eq:fiducial}
\end{align}

The time evolution of the particle production parameter  $\xi$ and of $m_Q$ is shown in Fig.~\ref{fig:fiducial}. Dashed lines denote the evolution of the parameters obtained neglecting scalar backreaction whilst solid lines denote the evolution of the full system. When backreaction becomes strong, around $N\simeq 32$, the gauge field $Q$ quickly decays to zero compared to the evolution in the absence of strong backreaction. This numerical result is in full agreement with the qualitative understanding detailed above. The reason for the rapid decay is the fast emergence of a quartic contribution to the potential, which forces the rapid decay of the gauge field background. 

The axion particle production parameter $\xi$ exhibits a brief but explosive increase before the axion reaches the bottom of the potential, where it decays as usual. During this ``spike-like" increase, the particle production parameter attains very large values that could, in principle, spell trouble if tensor fluctuations are overproduced. We observe, however, that such overproduction never occurs since these large values of $\xi$ are attained only for a very limited time interval, much shorter than one e-fold. 

\begin{figure}[t]
  \centering
  \includegraphics[width=\columnwidth]{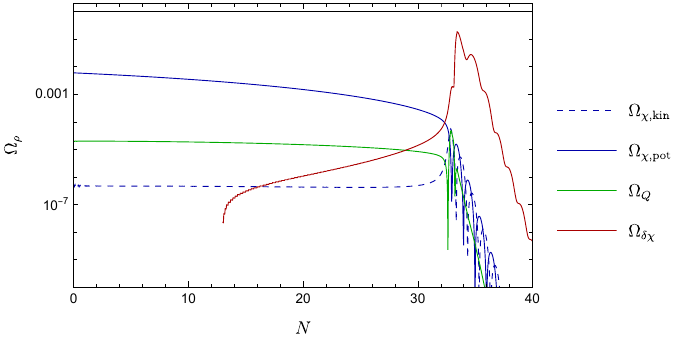}
  \caption{Comparison of the fractional energy density contribution of each component of our system. It is possible to observe that the inflaton energy (black horizontal line) remains the dominant component throughout the time span of our simulation.}
  \label{fig:Energies}
\end{figure}
In closing this section, we plot the relative contributions to the energy densities in Fig.~\ref{fig:Energies}. Other than the kinetic and potential energy of the axion, plotted separately, we also plot the energy of the gauge field background, which can be read off from the 1st Friedmann equation,
\begin{equation}
    \rho_{Q} = \frac{3}{2}(\dot{Q} + H Q)^2+ \frac{3}{2}g^2 Q^4\,.
\end{equation}
Additionally,one can derive the energy density of the axion scalar fluctuations by computing the mixed $00$ component of the energy momentum of the system,
\begin{equation}
\begin{split}
\rho_{\delta\chi} &= \frac{1}{2}\int \frac{d^3k}{(2\pi)^3}
\Bigg\{ |\dot{{\delta\chi}} |^2 \\
&\qquad + \biggl[ \Bigl(\frac{k}{a}\Bigr)^2 + U''(\chi)
+ \frac{g^2 k^2 \lambda^2 Q^4}{f^2\bigl(k^2 + 2 g^2 a^2 Q^2\bigr)}
\biggr]\,|\delta\chi|^2\Bigg\}\,.
\end{split}
\end{equation}
In order to facilitate the comparison of these energy densities with the inflaton energy, we plot fractional energy densities instead, defined as $\Omega_\rho\equiv\rho_i/(3H^2 M_p^2)$. Since throughout this work it is assumed that the inflaton dominates the energy budget of the universe, the inflaton density will be $\Omega_{\phi}=1$ by definition, and is shown as a black horizontal line in Fig.~\ref{fig:fiducial}.

The main takeaway from the fractional energy density comparison is that, once the axion-gauge system enters the tachyonic band ($m_Q<\sqrt{2}$), the energy density of scalar fluctuations increases progressively until very close to the decay of the axion-gauge system. Shortly before the axion-gauge system starts to decay, there is explosive production of scalar fluctuations which, however lose their support once the axion-gauge field backgrounds start redshifting rapidly.\footnote{In calculating the energy density of fluctuations, we are accounting for the contribution of those wavenumbers that experience tachyonic growth. We use the same UV and IR cutoffs employed for the backreaction integrals in Sec.~\ref{sec:backreaction}.} 

This mechanism cannot be used to produce PBHs or SIGWs since the tachyonically produced scalar fluctuations become massive and decay rapidly at superhorizon scales. As a result, these fluctuations are very small during horizon re-entry after inflation which is when PBHs and SIGWs are expected to be sourced. In the next section, we will explore the direct production of GWs due to the spike in particle production parameter $\xi$.
%
%%%%%%%%%%%%%%%%%%%%%%%%%%%%%%%%%%%%%%%%%%%%%%%%%%%%%%%%
\section{Phenomenology}
\label{sec:pheno}
%%%%%%%%%%%%%%%%%%%%%%%%%%%%%%%%%%%%%%%%%%%%%%%%%%%%%%%%
%
We now turn our attention to the phenomenology that results from the strong backreaction effects analyzed in the previous section. Although scalar fluctuations decay rapidly at superhorizon scales, the abrupt $\xi$ spike (see Fig.~\ref{fig:fiducial})  directly excites tensor modes, yielding a peaked stochastic primordial GW background. For the fiducial set of parameter values in Eq.~(\ref{eq:fiducial}), the peak is at interferometer scales.

The production of tensor modes via tachyonic instability has already been studied e.g. in \cite{Dimastrogiovanni:2016fuu} in the so-called CNI attractor regime, for which the two key parameters are related by $\xi=m_Q+1/m_Q$. This relation is of course no longer valid in the strong backreaction regime.  We therefore solve the equations of motion for the tensor fluctuations numerically without making any assumptions about  $\xi(m_Q)$. There are four degrees of freedom in the tensor sector in this model, two of which correspond to fluctuations of the metric and two of the gauge field. We decompose the field fluctuations as follows

\begin{equation}
    \delta g_{i j }= a^2 h_{ij},\qquad \delta A^a_i=a \,t^a_i
\end{equation}

where both $h_{ij}$ as well as $t^a_i$ are by construction transverse and traceless in the two indices. The presence of the homogeneous and isotropic gauge field background $Q\, \delta^a_i$ allows for these fluctuations to couple at the linearized level. We define the circular polarization according to the standard convention, $h_{R,L}\equiv\left(h_{+}\pm h_\times\right)/\sqrt{2}$ and $t_{R,L}\equiv\left(t_{+}\pm t_\times\right)/\sqrt{2}$.

After converting to Fourier space, we identify the normalized fields that yield a canonically normalized quadratic Lagrangian for the system of fluctuations
\begin{equation}
    h_{R, L} = \frac{\sqrt{2}}{M_{Pl}a} \psi_{R, L}, \qquad t_{R, L} = \frac{1}{\sqrt{2}a}T_{R, L}\;.
\end{equation}
Variation of this quadratic Lagrangian results in the equations of motion
\begin{widetext}
\begin{align}
\ddot{T}_{L,R} +H\,\dot{T}_{L,R} + & \left\{\frac{k^2}{a^2} + 2 H^2 \left[ m_Q\xi \pm \frac{k}{a H}(m_Q+\xi) \right] \right\} T_{L,R} =\nonumber\\
&+2 H\sqrt{\epsilon_{E}}\,\dot{\psi}_{L,R} - 2 H^2 \left[ \sqrt{\epsilon_{B}}\left(m_Q-2\xi \mp \frac{k}{a H}\right) + \sqrt{\epsilon_{E}}\right]\psi_{L,R} \label{eq:TRL}\\
\ddot{\psi}_{L,R} +H \dot{\psi}_{L,R}+& \left(\frac{k^{2}}{a^2} - 2 H^2\right)\psi_{L,R} =- 2 H \sqrt{\epsilon_{E}}\,\dot{T}_{L,R} + 2 H^2 \sqrt{\epsilon_{B}}\,\left(m_Q \mp \frac{k}{a H}\right)\,T_{L,R} \;.\label{eq:psiRL}
\end{align}
\end{widetext}
The canonical fields are initialized with the usual adiabatic vacuum $\psi_{{\rm in},R,L}=T_{{\rm in},R,L}\simeq \frac{{\rm e}^{-i k\tau}}{\sqrt{2k}}$.

These equations make explicit how the transient enhancements in  $\xi$ feed into the effective mass and mixing terms for the tensor sector. The parity breaking inherited from the Chern-Simons coupling, encoded in the $\pm$ terms, leads to a helicity-asymmetric amplification whose duration is controlled by the short-lived strong backreaction regime. We solve Eqs. (\ref{eq:TRL}) and (\ref{eq:psiRL}) numerically for a set of $k$-modes, with the evolution of the background quantities derived from the system that includes strong scalar backreaction. 

To make a connection with observations, we compute the present-day GW energy density. The GW energy density power spectrum in units of the critical density is given by \cite{Caprini:2018mtu}
\begin{equation}
\Omega_{\rm GW}(k) = \frac{3}{128}\Omega_{\rm rad}P^{\rm tot}_h(k)\left[\frac{1}{2}\left(\frac{k_{\rm eq}}{k}\right)^2 + \frac{16}{9}\right],
\label{eq:omega_gw}
\end{equation}
where $\Omega_{\rm rad} \simeq h^{-2} \times 2.47 \times 10^{-5}$ is the current radiation density parameter and $k_{\rm eq} \simeq 1.3 \times 10^{-2} \, \text{Mpc}^{-1}$ denotes the comoving wave number corresponding to horizon entry at matter-radiation equality. The dimensionless Hubble parameter $h$ is defined through $H_0 = 100h \, \text{km} \, \text{s}^{-1} \, \text{Mpc}^{-1}$. The total tensor power spectrum $P_h^{\text{tot}}(k)$ includes both the vacuum contribution and the sourced component obtained from the numerical evolution of Eqs.~(\ref{eq:TRL}) and (\ref{eq:psiRL}).

\begin{figure}[t]
  \centering
\includegraphics[width=\columnwidth]{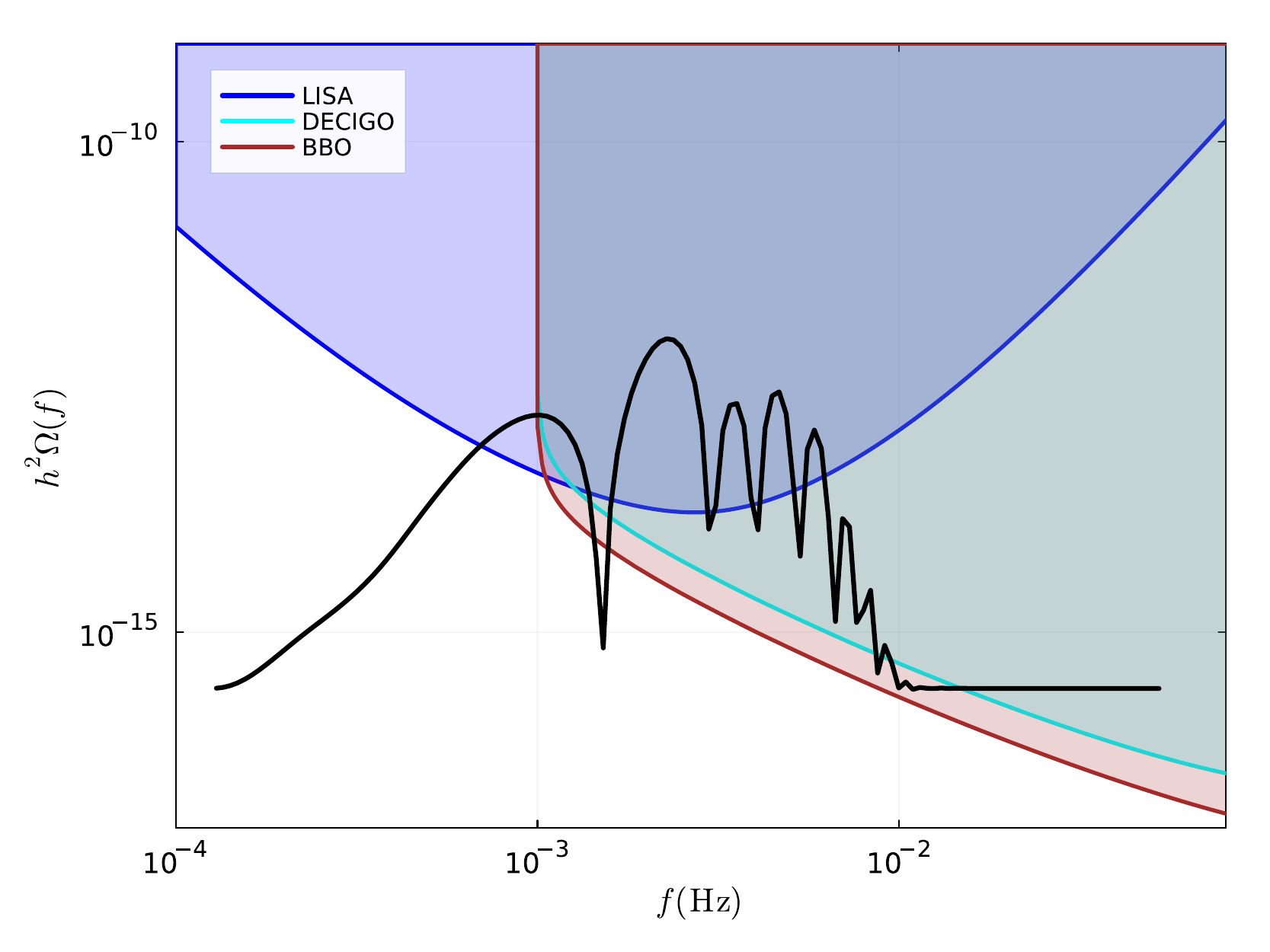}
  \caption{Gravitational Wave energy density. We use for our numerical simulation the fiducial values listed in Eq.~\ref{eq:fiducial}.}
  \label{fig:GW}
\end{figure}

For the benchmark parameters choice of Eq.~\eqref{eq:fiducial}, the resulting GW spectrum $\Omega_\text{GW}(f)$, shown in Fig.~\ref{fig:GW}, exhibits a localized peak at frequencies corresponding to modes exiting the horizon around the time of the $\xi$ spike. The amplitude of this peak is smaller than that achieved in the strong tensor backreaction regime of axion–SU(2) setups, where the gauge tensor modes themselves experience a prolonged tachyonic growth. It is nevertheless sufficiently large to be detectable by upcoming space-based interferometers, \cite{Dimastrogiovanni:2024xvc}. For our fiducial set of values in SCNI, the peak falls within the expected sensitivity band of LISA and would also be detected by futuristic missions targeting a similar frequency range such as like DECIGO and BBO. Additionally, this mechanism opens the possibility to obtain an observable GW signal even for a small value of the Chern-Simons coupling which is easier to ember within UV completions \cite{Bagherian:2022mau}.

The position of the peak of the GW signal depends on the choice of initial conditions for the axion background value at early times, when the CMB modes crossed the horizon.  From the string axiverse \cite{Arvanitaki:2009fg, Obata:2014loa, Cicoli:2013ana, Preskill:1982cy, Abbott:1982af, Sikivie:2006ni, Harari:1987ht, Dimastrogiovanni:2023juq, Acharya:2010zx, Cicoli:2012sz, Demirtas:2018akl, Demirtas:2021gsq, DAmico:2021vka, DAmico:2021fhz} perspective, one would expect the presence of multiple axions with different masses, decay constants, couplings and initial conditions  (see Ref.~\cite{Dimastrogiovanni:2023juq}) so that a whole ``gravitational wave forest\footnote{Depending on the nature of the distribution, one expects only a fraction of this ``forest'' to be detectable, the one corresponding to the appropriate choice of parameters.}'' would be in place.

An important feature worth stressing about the scenario we studied is that the GW signal is generated in a regime where scalar perturbations are always kept under control. The same tachyonic instability that triggers strong scalar backreaction is responsible for the rapid decay of the axion–gauge background, ensuring that scalar modes become heavy and redshift away rather than forming PBHs or large scalar-induced GW backgrounds. As a result, the regime we studied offers a robust, predictive, link between a well-defined strong backreaction regime in the spectator sector and an observationally testable GW signature.Further exploration of the parameter space, or studies analogous to the recent \cite{Michelotti:2024bbc,Ishiwata:2025wmo}, and ultimately a fully non-perturbative analysis via lattice simulations (as has been extensively done in the case of Abelian gauge fields \cite{Caravano:2022epk,Figueroa:2023oxc,Caravano:2024xsb,Figueroa:2024rkr,Sharma:2024nfu,Lizarraga:2025aiw,Iarygina:2025ncl}), will allow us to go beyond the regime studied here. We leave this for future investigation.
%
%%%%%%%%%%%%%%%%%%%%%%%%%%%%%%%%%%%%%%%%%%%%%%%%%%%%%%%%
\section{Conclusions}
\label{sec:conclusions}
%%%%%%%%%%%%%%%%%%%%%%%%%%%%%%%%%%%%%%%%%%%%%%%%%%%%%%%%
%
We performed a concrete numerical analysis of the dynamics of strong scalar backreaction in models of inflation featuring axion-SU(2) couplings. This and similar scenarios are known to feature a tachyonic instability in the scalar fluctuations. Tachyonically enhanced modes acquire large occupation numbers, which can backreact and alter the background dynamics. While our perturbative treatment provides the first reliable account of scalar backreaction dynamics in these models, a fully non-perturbative analysis via lattice simulations will shed light on any remaining subtleties.

With this caveat in mind, our results reveal a novel strong scalar backreaction regime whose qualitative features are primarily controlled by the value of the axion decay constant $f$. We identified a range of values for $f$ within which our numerical analysis is under control and described how the effective potential of the gauge field background is modified as a result of backreaction. This modification typically takes the form of an additional quartic potential term, which leads to the rapid decay of the gauge field background. In turn, this leads to a sharp and short-lived peak in the particle production parameter $\xi$ associated with the axion. We found that the spike in  $\xi$ results in a ``bumpy" GW spectrum which features relatively rapid oscillations in frequency space. The sourced GWs are almost entirely chiral as is usually the case in axion-gauge field inflation. The position of the peak of the spectrum is controlled by the initial conditions. For our fiducial set of parameter values, the GW signal peak is in the LISA frequency band and of sufficient amplitude to grant detection. This is particularly intriguing from the point of view of the string axiverse, given that the latter posits a host of axion sectors whose presence could be revealed by their gravitational wave imprints.

%%%%%%%%%%%%%%%%%%%%%%%%%%%%%%%%%%%%%%%%%%%%%%%%%%%%%%%%
\begin{acknowledgments}
It's a pleasure to thank Cristóbal Zenteno Gatica for helpful discussions. MF and AP acknowledge the “Consolidación Investigadora” grant CNS2022-135590. The work of MF and AP is partially supported by the Spanish Research Agency (Agencia Estatal de Investigación)
through the Grant IFT Centro de Excelencia Severo Ochoa No CEX2020-001007-S, funded by
MCIN/AEI/10.13039/501100011033. MF acknowledges support from the “Ramón y Cajal” grant
RYC2021-033786-I. 
The work of MC is supported by Ministero dell’Universit\`a e della Ricerca (MUR), PRIN2022 program (Grant PANTHEON 2022E2J4RK) Italy, and 
by the Research grant TAsP (Theoretical Astroparticle Physics) funded by Istituto Nazionale di Fisica Nucleare (INFN). Part of this work was carried out during the 2025 ``The Dawn of Gravitational Wave Cosmology'' workshop, supported by the Fundaci\'{o}n Ram\'{o}n Areces and hosted by the ``Centro de Ciencias de Benasque Pedro Pascual''. We thank both the CCBPP and the Fundaci\'{o}n Areces for creating a stimulating and very productive environment for research.
\end{acknowledgments}

\appendix

%%%%%%%%%%%%%%%%%%%%%%%%%%%%%%%%%%%%%%%%%%%%%%%%%%%%%%%%
\section{Equations of motion for canonical scalar fluctuations}
\label{app:eoms}
%%%%%%%%%%%%%%%%%%%%%%%%%%%%%%%%%%%%%%%%%%%%%%%%%%%%%%%%

The equations of motion for the canonically normalized scalar fluctuations take the form
\newcommand{\XX}{\hat{X}}
\newcommand{\PS}{\hat{\Phi}}
\newcommand{\ZZ}{\hat{Z}}
\newcommand{\dem}{k^2 + 2  m_Q^2\,a^2 H^2 }
\begin{widetext}
\begin{align}
    \ddot{\XX} & + H \dot{\XX} + 
    \left[
    -2 + \epsilon_H + 3 \eta_\chi + 
    \frac{ k^2\left(k^2 +  m_Q^2 (2 + \Lambda)\,a^2 H^2 \right)}{a^2 H^2\left(k^2 + 2  m_Q^2\,a^2 H^2\right)} \right] H^2\XX+\frac{\sqrt{2} m_Q^2 \Lambda\, aH}{\sqrt{k^2 + 2  m_Q^2\,a^2 H^2}} H\dot{\PS} \nonumber\\
    &+    
   \frac{m_Q\Lambda\left(k^4 + 3m_Q^2\, k^2  a^2H^2 + 4 m_Q^4\,  a^4H^4  \right) }{aH \left(k^2 + 2 m_Q^2\,a^2 H^2\right)^{3/2}} \sqrt{\frac{2\epsilon_E}{\epsilon_B}}H^2\PS - \sqrt{2} m_Q \Lambda\, H \dot{\ZZ} -
   2 m_Q^2\Lambda  \sqrt{\frac{2\epsilon_E}{\epsilon_B}}\,H^2 \ZZ  = 0,
   \label{eq:scalarpertX}\\
    \ddot{\PS} &+ H \dot{\PS} +  
    \left[ 
    \frac{6 m_Q^4\,k^2\, a^2 H^2}{ \left(\dem\right)^2}\frac{\epsilon_E}{\epsilon_B} +
    \frac{k^4 + 2 m_Q(3m_Q - \xi)k^2\, a^2H^2 + 4m_Q^4\, a^4 H^4}{\left(\dem\right)\,a^2H^2}
    \right]H^2 \PS \nonumber\\
    &- \frac{2 (m_Q - \xi)\sqrt{\dem}}{a H}\, H^2\ZZ - \frac{\sqrt{2} m_Q^2 \Lambda\, a H}{\sqrt{\dem}}H\dot{\XX} 
    \nonumber\\
    &
    +\frac{\sqrt{2} m^2_Q \Lambda\, a H}{\sqrt{\dem}}\left[1+
    \frac{ k^4}{m_Q\left(\dem\right)\, a^2 H^2}\sqrt{\frac{\epsilon_E}{\epsilon_B}} 
    \right] H^2 \XX = 0,
    \label{eq:scalarpertPsi}\\
    \ddot{\ZZ} &+ H \dot{ \ZZ} +  \left[\frac{k^2}{a^2H^2} + 2m_Q \left(m_Q - 2 \xi\right)   \right]H^2 \ZZ 
    + \sqrt{2} m_Q \Lambda\, H \dot{\XX} - \sqrt{2} m_Q \Lambda\, H^2 \XX
    \nonumber \\
    & -
    \frac{2 \left(m_Q - \xi\right)\sqrt{\dem} }{aH}H^2\PS = 0.
    \label{eq:scalarpertZ}
\end{align}
\end{widetext}
where we defined the slow roll parameters
\begin{align}
    \epsilon_\chi\equiv \frac{\dot{\chi}^2}{2M_p^2 H^2}\,,&\;\;\epsilon_B\equiv \frac{g^2Q^4}{M_p^2 H^2}\,,\;\;\epsilon_E\equiv \frac{\left(\dot{Q}+H Q\right)^2}{M_p^2 H^2}.
\label{eq:slow}
\end{align}
and parameter $\eta_\chi\equiv\frac{V_{\chi\chi}}{3 H^2}$ and the dots represent the derivatives with respect to the cosmic time $t$.

%%%%%%%%%%%%%%%%%%%%%%%%%%%%%%%%%%%%%%%%%%%%%%%%%%%%%%%%
\section{Initial conditions for fluctuations and numerical solutions}
\label{app:initial}
%%%%%%%%%%%%%%%%%%%%%%%%%%%%%%%%%%%%%%%%%%%%%%%%%%%%%%%%
%
The system of equations in App.\ref{app:eoms} is a linear system which couples the canonical axion fluctuation $\hat{X}$ with the dynamical, canonical, scalar fluctuations of the gauge field $\hat{\Phi}$ and $\hat{Z}$. In order to set the initial conditions, we again perform an expansion deep inside the horizon and assume $\Lambda\gg\sqrt{2}$. The initial conditions have been derived in Appendix A of \cite{Dimastrogiovanni:2024xvc} and we merely reiterate them here for completeness. 
\begin{align}
    \hat{X}_{\rm in}=&\frac{\sqrt{1+m_Q^2}}{\sqrt{2k}}{\rm e}^{i \left(x-x_{\rm in}\right)}\left(\frac{x_{\rm in}}{x}\right)^{i\sqrt{\frac{1+m_Q^2}{2}}\Lambda}\;,\nonumber\\
    \hat{\Phi}_{\rm in}=&-\frac{1}{\sqrt{2k}}{\rm e}^{i \left(x-x_{\rm in}\right)}\left(\frac{x_{\rm in}}{x}\right)^{i\sqrt{\frac{1+m_Q^2}{2}}\Lambda}\;,\nonumber\\
    \hat{Z}_{\rm in}=&i\frac{m_Q}{\sqrt{2k}}{\rm e}^{i \left(x-x_{\rm in}\right)}\left(\frac{x_{\rm in}}{x}\right)^{i\sqrt{\frac{1+m_Q^2}{2}}\Lambda}\;.\nonumber\\
\end{align}

The numerical strategy for solving these equations is mostly inspired by the numerical work of the Axion-U(1) scenario employed in \cite{Garcia-Bellido:2023ser}. We solve the background equations for the axion (\ref{eq:chieom}) and gauge field (\ref{eq:eoms}) background in tandem with the equations for fluctuations (\ref{eq:scalarpertX}),(\ref{eq:scalarpertPsi}),(\ref{eq:scalarpertZ}) for $i_{\rm max}$ different wave numbers equidistant in $\log k$ space. The backreaction term is reconstructed at every moment in time using the trapezoidal rule with UV and IR cutoffs as explained in the main text \ref{sec:backreaction}.

We work with the number of e-folds as our time variable throughout our numerical implementation, and we transform our equations for fluctuations to auxiliary variables defined as
\begin{align}
    \tilde{X}\equiv&\frac{\hat{X}}{\hat{X}_{\rm in}}\;,\nonumber\\
    \tilde{\Phi}\equiv&\frac{\hat{\Phi}}{\hat{\Phi}_{\rm in}}\;,\nonumber\\
    \tilde{Z}\equiv&\frac{\hat{Z}}{\hat{Z}_{\rm in}}\;.\nonumber\\
\end{align}
This transformation requires recasting the equations of motion for the fluctuations into equations with respect to the tilde variables. Even though this results in a more complex set of equations, the transformation above is highly advantageous as it effectively cancels out the fast oscillations inside the horizon, making the system of equations faster to solve, while simultaneously changing the initial conditions into ones and zeros for the auxiliary fields and their derivatives. 

\begin{figure}[t]
  \centering
\includegraphics[width=\columnwidth]{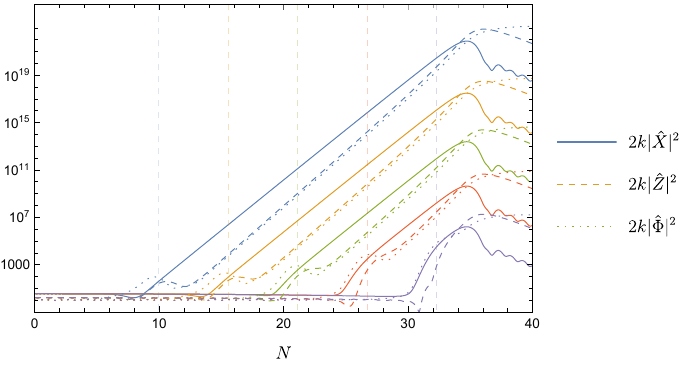}
  \caption{Comparison of the time evolution of the canonical fluctuations of the axion field (solid) and of the gauge field (dashed and dotted). Different colors indicate fluctuations of different wave numbers and vertical dashed lines show the horizon crossing time for each of them.}
  \label{fig:mode-functions}
\end{figure}

As an example of the typical evolution of fluctuations in this scenario we plot the canonically normalized mode functions in Fig.\ref{fig:mode-functions}. We have selected fluctuations of five different wave numbers which cross the horizon at a time indicated by the vertical semi-transparent dashed lines. The solid lines indicate the axion fluctuation, while the dashed and dotted lines are the gauge field fluctuations. It is easy to observe that the tachyonic instability is active inside the horizon, and as a result, the occupation number of these fluctuations increases rapidly. Right after horizon crossing the fluctuations of the gauge field decay, leaving the axion fluctuation as the dominant one. This justifies our neglect of the gauge field fluctuations for the purpose of estimating the backreaction, as explained in the main text. 

\bibliography{Backreaction}% Produces the bibliography via BibTeX.

\end{document}